\documentclass[11pt,letterpaper]{article}

\usepackage{definitions}
\usepackage[retainorgcmds]{IEEEtrantools}
\usepackage{amssymb}
\usepackage{amsmath}
\usepackage{multirow}

\font\ro=cmsy10                          % font with rope
\def\kcr{{\hbox{\ro \char'170}}}                % right-handed rope
\def\ktl{{\hbox{\ro \char'170}}}        % top end for left-handed rope
\def\ktr{{\hbox{\ro \char'170}}}        % " right
\def\kbl{{\hbox{\ro \char'170}}}        % " bottom left
\def\kbr{{\hbox{\ro \char'170}}}        % " right
\parskip=\medskipamount                 % space between paragraphs (LaTeX)
\thicklines                         % thick straight lines for pictures (LaTeX)

\def\border{                                            % border
        \setlength{\unitlength}{1mm}
        \newcount\xco
        \newcount\yco
        \xco=-21
        \yco=12
        \begin{picture}(140,0)
        \put(\xco,\yco){$\ktl$}
        \advance\yco by-1
        {\loop
        \put(\xco,\yco){$\kcr$}
        \advance\yco by-2
        \ifnum\yco>-240
        \repeat
        \put(\xco,\yco){$\kbl$}}
        \xco=158
        \yco=12
        \put(\xco,\yco){$\ktr$}
        \advance\yco by-1
        {\loop
        \put(\xco,\yco){$\kcr$}
        \advance\yco by-2
        \ifnum\yco>-240
        \repeat
        \put(\xco,\yco){$\kbr$}}
        \put(-20,13){\tiny **University of Maryland * Center for String and
         Particle  Theory* Physics Department***University of Maryland *Center
        for String and Particle  Theory** }
        \put(-20,-241.5){\tiny **University of Maryland * Center for String and
         Particle  Theory* Physics Department***University of Maryland *Center
        for String and Particle  Theory** }
        \end{picture}
        \par\vskip-8mm}

\def\headpic{                                           % UM heading
        \indent
        \setlength{\unitlength}{.4mm}
        \thinlines
        \par
        \begin{picture}(29,16)
        \put(165,16){\line(1,0){4}}
        \put(170,16){\line(1,0){4}}
        \put(180,16){\line(1,0){4}}
        \put(175,0){\line(1,0){4}}
        \put(180,0){\line(1,0){4}}
        \put(185,0){\line(1,0){4}}
        \put(169,0){\line(0,1){16}}
        \put(170,0){\line(0,1){16}}
        \put(179,0){\line(0,1){16}}
        \put(180,0){\line(0,1){16}}
        \put(184,0){\line(0,1){16}}
        \put(185,0){\line(0,1){16}}
        \put(169,16){\oval(8,32)[bl]}
        \put(170,16){\oval(8,32)[br]}
        \put(179,0){\oval(8,32)[tl]}
        \put(185,0){\oval(8,32)[tr]}
        \end{picture}
        \par\vskip-6.5mm
        \thicklines}
\def\endtitle{\end{quotation}\newpage}                  % end title page

\begin{document}

\border\headpic {\hbox to\hsize{October 2013 \hfill
{UMDEPP-013-015}}}
\par
{$~$ \hfill
%{hep-th/xxxx.xxxx}
}
\par

\setlength{\oddsidemargin}{0.3in}
\setlength{\evensidemargin}{-0.3in}
\begin{center}
\vglue .10in
{\large\bf On 4$D$, ${\cal N} = 1$ Massless Gauge Superfields of\\
Higher Superspin: Integer Case
 \footnote
{Supported in part  by National Science Foundation Grant
PHY-09-68854.}\  }
\\[.5in]

S.\, James Gates, Jr.\footnote{gatess@wam.umd.edu}
and
Konstantinos Koutrolikos\footnote{koutrol@umd.edu}
\\[0.2in]

{\it Center for String and Particle Theory\\
Department of Physics, University of Maryland\\
College Park, MD 20742-4111 USA}\\[1.8in]

{\bf ABSTRACT}\\[.01in]
\end{center}
\begin{quotation}
{We present an alternative method of exploring the component 
structure of an integer super-helicity $Y=s$ (for any integer $s$) 
irreducible representation of the Super-Poincar\'{e} group. We 
use it to derive the component action and the SUSY transformation 
laws. The effectiveness of this approach is based on the equations 
of motion and their properties, like Bianchi identities. These 
equations are generated by the superspace action when it is 
expressed in terms of prepotentials.  For that reason we reproduce 
the superspace action for integer superspin, using unconstrained 
superfields. The appropriate, to use, superfields are dictated by 
the representation theory of the group and the requirement that 
there is a smooth limit between the massive and massless case.
}

%${~~~}$ \newline
%PACS: 04.65.+e
\endtitle

\section{Introduction}
~~ Higher spin field theory has a very rich history driving the 
developments of modern theoretical physics and after many 
decades still remains a very active subject. It started with Dirac 
\cite{Dirac} trying to generalize his celebrated spin-$\frac{1}{2}$ 
equation. His comment in that paper ``the underlying theory is 
of considerable interest'' still resonates. After the classical work 
by Fierz and Pauli \cite{FP} there was an increasing  number 
of papers formulating the theory of a massive arbitrary spin in 
four dimensions \cite{SH1,SH2} as well as developments for the 
massless arbitrary helicities using the `principle' of gauge 
invariance \cite{FdL,F&F}. Since then there has been tremendous 
progress with generalizations of these results regarding irreducible 
representations of the little group in $D$-dimensions \cite{V}, 
derivations of the massive theories by means of dimensional 
reduction of the massless theories in $D+1$-dimensions
\cite{BB1}, St\"{u}ckelberg formulations
\cite{R&S}, BRST \cite{BB2}, quantization 
and many other things.

The discussion of arbitrary spin gauge fields in the context of
simple symmetry in four dimensions parallels this development
of the general discussion.  At the level of component fields
this was initiated by Curtright \cite{Curt}, followed by the
superfield discussion at the level of on-shell equations of
motion \cite{Sspace}, and finally followed by the off-shell
discussions in the work of Kuzenko, et.\ al. \cite{Off1,Off2}.
These pioneering works on higher spin 4$D$, $\cal N$ = 1
supermultiplets have also led to the creation of a growing
literature \cite{HyRspnLIT} on the subject.

A current generator of interest about higher spin theories has 
been generated by string theory as it low-energy approximation
leads to consideration of fields of unbounded spins since the 
spectrum of string and superstring theory includes an infinite 
tower of massive spin states. Therefore a limit must exist where 
(super)string theory is formulated as a field theory of interacting 
spins. That points to the interesting direction of extending all 
previous results to include supersymmetry. The tool to build 
4$D$, ${\cal N}$ = 1 manifestly SUSY invariant theories is superspace 
and the usage of superfields.

For the massless case such a construction exists \cite{Off1,Off2}. 
The theories presented in these works, were initially described
in terms of constrained superfields. The purpose of the differential 
constraints is to achieve gauge invariance. As they comment in 
their work these constraints can easily be solved in terms of prepotentials. 
These prepotentials can play a role in the formulation of massive 
superspin theories and maybe even spin interaction theories.  
In a subsequent work \cite{Kuzp}, these unconstrained prepotentials
were introduced and used to show that the works of \cite{Codi}
occur by applying a transformation to the original formulations.

In this current work (and an accompanying one \cite{IntSpin}) we 
would like to show how representation theory of the Super-Poincar\'{e} 
group makes these prepotential variables building blocks for 
massive and massless theories and then use them to reproduce 
the realizations of irreducible representations with arbitrary super-helicity.

In the previous works, when discussion about the component field 
spectrum of the theories was given, it was based on $\theta$-expansion 
of the superfields in the Wess-Zumino gauge.  This implied that by 
using that ansatz for the components and the usual rules of projection, 
the component action and the SUSY-transformation laws can be derived. 

This process is straightforward but cumbersome. For this reason we 
exploit an alternative efficient way of defining components, using the 
superfield equations of motion. The action itself, with the help of the 
Bianchi identities, will guide us to efficient definitions of the components, 
derive the component action  and the SUSY-transformation laws.  This
approach builds naturally on \cite{GrvTn} for the study of the component 
structure of super-helicity $Y=1$ and discussions \cite{5D} 
on old-minimal supergravity. 

However there is a key difference with both of these.  The first one used 
the superfield strength as a guide for the definition of the components. 
This approach can not be generalized for the arbitrary super-helicity 
because of the mass dimensionality of the superfield strength is proportional 
to super-helicity. In the second paper components were defined without 
finding the component action and SUSY-transformation laws. We will do 
both of these for the arbitrary integer super-helicity case

In this follow, we focus on arbitrary integer super-helicity irreducible 
representation of the $4D$, $\cal N$ = 1 Super-Poincar\'{e} group. A 
discussion for the half-integer super-helicities will presented in a 
following letter. The presentation is organized as follows: In section 
\ref{Irreps} we briefly review the representation theory of the little 
group of the 4$D$, $\cal N$ = 1 Super-Poincar\'{e} group, following 
\cite{BB2}. This discussion will illuminate the proper superfields one 
should use in order to construct the desired representations. In section 
\ref{Redndcy} we focus on the massless integer super-helicity case 
and illustrate how the principle of gauge invariance emerges from the 
requirement to have a smooth transition between massive and massless 
theories. In section \ref{Action} we find the superspace action of the theory 
and prove that it describes the desired super-helicity. The last section 
\ref{Projection} is a discussion about the off-shell component structure 
of the theory. We present a self-contained method of defining the 
components, find the component action and give explicit expressions 
for the SUSY-transformation laws.  The main new results in this (and a 
companion) work involve the derivation of a complete
component-level description that involves {\em {no}} explicit $\theta$-expansion
of superfields.  The conventions used are the ones of \cite{5D}.

\section{Irreducible Representations}
\label{Irreps}
~~ As is well known the Super-Poincar\'{e} group has two 
Casimir operators that label the irreducible representations. 
The first one is the mass and the other one is a supersymmetric 
extension of the Poincar\'{e} Spin operator.

\subsection{Massive Case}
For the massive case the second casimir operator takes the form
\be
C_2=\frac{W^2}{m^2}+\left(\frac{3}{4}+\lambda\right)P_{(o)}
\ee
where $W^2$ is the Poincar\'{e} Spin operator, $P_{(o)}$ is a 
projection operator and the parameter $\lambda$ satisfies the 
equation
\be
\lambda^2+\lambda=\frac{W^2}{m^2}
\ee
In order to diagonalize $C_2$ we want to diagonalize both 
$W^2,\ P_{(o)}$. The superfield $\Phi_{\a(n)\ad(m)}$ which
does that and describes the highest possible superspin
representation
\bea{l}
C_2\Phi_{\a(n)\ad(m)}=Y(Y+1)\Phi_{\a(n)\ad(m)}~,~Y=
\frac{n+m+1}{2}\IEEEyesnumber
\eea
has to satisfy the following constraints:
\bea{l}
\text{symmetrized dotted and undotted indices}\\
\D^2\Phi_{\a(n)\ad(m)}=0\\
\Dd^2\Phi_{\a(n)\ad(m)}=0\\
\D^{\g}\Phi_{\g\a(n-1)\ad(m)}=0\IEEEyesnumber\\
\pa^{\g\gd}\Phi_{\g\a(n-1)\gd\ad(m-1)}=0\\
\Box\Phi_{\a(n)\ad(m)}=m^2\Phi_{\a(n)\ad(m)}
\eea
All these can be satisfied if
\be
\Phi_{\a(n)\ad(m)}\sim\D^{\g}W_{\a(n)\g\ad(m)}~,~W_{\a(n+
1)\ad(m)}\sim\Dd^2\D_{(\a_{n+1}}\Phi_{\a(n))\ad(m)}\\
\ee
with
\bea{l}
\Dd_{\bd}W_{\g\a(m)\ad(n)}=0,\ \text{chiral}\\
\pa^{\b\bd}W_{\b\a(m)\bd\ad(n-1)}=0\IEEEyesnumber\\
\Box W_{\a(m+1)\ad(n)}=m^2W_{\a(m+1)\ad(n)}
\eea
and the spin content of this supermultiplet is $j=Y+1/2,\ Y,\ Y,\ Y-1/2$.

Therefore the superfield that describes a superspin $Y$ system, 
has index structure such that $n+m=2Y-1$ where $n,m$ are 
integers. This Diophintine equation has a finite number of solutions 
for $(n,m)$ pairs, but the corresponding superfields are all related 
because we can use the $\pa_{\b\bd}$ operator to convert one 
kind of index to another. So we can pick one of them to represent 
the entire class.

One last comment has to be made about the reality of the representation. 
Under a hermitian conjugation a $(n,m)$ representation realized by a 
superfield like $\Phi_{\a(n)\ad(m)}$ goes to a $(m,n)$ representation, 
realized by $\bar{\Phi}_{\a(m)\ad(n)}$
\bea{c}
(n,m)^*\rightarrow(m,n) \begin{cases} \text{if}~m=n,~& (n,
n)^*\rightarrow (n,n)~\text{:reality} \\ \text{if}~m\neq n,~& (n,
m)^*\rightarrow (m,n)\neq (n,m)\\ {}&\text{to make real representations}\\
{}& \text{we need to consider} ~(n,m)\oplus (m,n) \end{cases}
\eea

At the superfield level this mapping can be done by the dimensionless 
operator $\Delta_{\a\ad}\equiv-i\frac{\pa_{\a\ad}}{\Box^{1/2}}$ which if 
used in repetition will convert all the undotted indices to dotted ones 
and vice versa.
\bea{c}
\bar{\Phi}_{\a(m)\ad(n)}=\Delta_{a_1}{}^{\gd_1}\dots\Delta_{
a_m}{}^{\gd_m}\Delta^{\g_1}{}_{\ad_1}\dots\Delta^{\g_1}{}_{
\ad_1}\Phi_{\g(n)\gd(m)}
\eea

For irreducible representations with $n=m$ (bosonic superfields)
 the reality condition becomes $\Phi_{\a(n)\ad(n)}=\bar{\Phi}_{\a(
 n)\ad(n)}$ and for fermionic superfields ($n=m+1$) the reality 
 condition is the Dirac equation $i\pa_{\a_n}{}^{\ad_n}\bar{\Phi
 }_{\a(n-1)\ad(n)}+m\Phi_{\a(n)\ad(n-1)}=0$.
 
\subsection{Massless Case}
For the masssless case, the supersymmetric analogue to the 
Pauli-Lubanski vector takes the form
\be
Z_{\g\gd}=W_{\g\gd}+\frac{1}{4}[\D_{\g},\Dd_{\gd}]
\ee
and our goal is to make it proportional to momentum. The 
superfield $F_{\a(n)\ad(m)}$ which does that and describes 
the highest super-helicity
\bea{l}
Z_{\g\gd}F_{\a(n)\ad(m)}=\left(Y+\frac{1}{4}\right)P_{\g\gd}
F_{\a(n)\ad(m)}~,~Y=\frac{n-m}{2}\IEEEyesnumber
\eea
must satisfy the following:
\bea{l}
\text{symmetrized dotted and undotted indices}\\
\Dd_{\gd}F_{\a(n)\ad(m)}=0,\ \text{chiral}\\
\D^{\b}F_{\b\a(n-1)\ad(m)}=0\IEEEyesnumber\\
\pa_{\g}{}^{\bd}F_{\a(n)\bd\ad(m-1)}=0
\eea
and the helicity content is $h=Y+1/2,\ Y$

So the superfield that describes a system with super-helicity $Y$, 
must have index structure such that $n-m=2Y$.  This Diophintine 
equation has infinite many solutions with an increasing number 
of indices. Nevertheless all of them can be generated by acting 
with $\pa_{\b\bd}$ on the superfield with the fewest indices 
$F_{\a(2Y)}$.

\section{Integer super-helicity,\ $Y=s$}
\label{Redndcy}
~~ The above discussion suggests that a theory of massive 
integer superspin $Y=s$ must be constructed in terms of a 
fermionic superfield $\Psi_{\a(s)\ad(s-1)}$ and there exists 
a chiral superfield $W_{\a(s+1)\ad(s-1)}\sim\Dd^2\D_{(\a_{
s+1}}\Psi_{\a(s))\ad(s-1)}$.

On the other hand the theory of massless integer super-helicity 
must be described in terms of a chiral superfield $F_{\a(2s)}$.

Now let us assume we have managed to develop the theory of 
massive integer superspin. We should be able to take the 
massless limit of that. It would be nice if such a limit leads to the 
theory of massless integer super-helicity (plus possibly other 
sectors that decouple).  But we showed that these two theories 
are described by different objects.  How can this be?  For 
something like that to happen we have to able to construct 
an object like $F_{\a(2s)}$ out of the remaining objects after 
the limit has been taken. Given the chirality property of $F$ 
and $W$ and their index structure we could guess a mapping 
that could do the trick.
\be
F_{\a(2s)}\sim\pa_{(\a_{2s}}{}^{\ad_s}\ldots\pa_{\a_{s+1}}
{}^{\ad_1}\Dd^2\D_{\a_{s+1}}\Psi_{\a(s))\ad(s-1)}
\ee

But there is a problem with this map. The problem is that $
F_{\a(2s)}$ which describes the system and carries the 
physical degrees of freedom seems to be defined in terms 
of another object $\Psi_{\a(s)\ad(s-1)}$. Also $F$ as defined 
above seems to have the on-shell degrees of freedom of 
$\Psi$ which is more than needed.  If this is going to work 
we have to find a way to 1) remove the physical (observable) 
status of $\Psi$ and 2) remove its extra degrees of freedom.

There is a mechanism that can do both at the same time. 
That is to introduce a redundancy.  We identify $\Psi_{\a(
s)\ad(s-1)}$ with $\Psi_{\a(s)\ad(s-1)}+R_{\a(s)\ad(s-1)}$ 
and instead of talking about $\Psi_{\a(s)\ad(s-1)}$ we talk 
about equivalence classes. $\Psi_{\a(s)\ad(s-1)}\sim\Psi_{
\a(s)\ad(s-1)}+R_{\a(s)\ad(s-1)}$.  This redundancy has 
to respect the physical - propagating degrees of freedom 
of $F$ and leave them unchanged. Hence
\be
\pa_{(\a_{2s}}{}^{\ad_s}\ldots\pa_{\a_{s+1}}{}^{\ad_1}
\Dd^2\D_{\a_s}\bar{R}_{\a(s-1))\ad(s)}=0
\ee
The most general solution to that is
\bea{l}
R_{\a(s)\ad(s-1)}=\frac{1}{s!}\D_{(\a_s}K_{\a(s-1))\ad(s-1)}
+\frac{1}{(s-1)!}\Dd_{(\ad_{s-1}}\Lambda_{\a(s)\ad(s-2))}
\IEEEyesnumber
\eea
where $K_{\a(s-1)\ad(s-1)},~\Lambda_{\a(s)\ad(s-2)}$ are 
completely unconstrained superfields. It is obvious that this 
redundancy will be the starting point for the gauge invariance 
story.

\section{The Superspace Action}
\label{Action}
~~Using the equivalency class characterized by $\Psi$ and 
redundancy $R$ we attempt to construct a superspace action 
that will describe the irreducible representation of integer 
super-helicity.  For that $\Psi$ must have mass dimensions 
$1/2$\footnote{it's highest spin component is a propagating 
fermion.} and the action must involve two covariant 
derivatives.\footnote{The action must be quadratic in $\Psi$ 
and dimensionless.}

The most general action is
\bea{ll}
S=\int d^8z&\ a_1\Psi^{\a(s)\ad(s-1)}\D^2\Psi_{\a(s)\ad(
s-1)}+c.c.\nonumber\\
&+a_2\Psi^{\a(s)\ad(s-1)}\Dd^2\Psi_{\a(s)\ad(s-1)}
+c.c.\nonumber\\
&+a_3\Psi^{\a(s)\ad(s-1)}\Dd^{\ad_s}\D_{\a_s}\bar{
\Psi}_{\a(s-1)\ad(s)}\nonumber\\
&+a_4\Psi^{\a(s)\ad(s-1)}\D_{\a_s}\Dd^{\ad_s}\bar{
\Psi}_{\a(s-1)\ad(s)}\nonumber
\eea
The goal is to find an action that respects the redundancy. 
That is the starting point for gauge invariance $\delta_{G}S=0$. 
The strategy to obtain this is to pick the free parameters in a 
special way. If this is not possible then we introduce auxiliary 
superfields, compensators and/or impose constraints on the 
parameters of the redundancy (gauge parameters). We also
assume it is reasonable to expect any compensators introduced, 
if necessary, will not introduce degrees of freedom with spin 
higher or equal than the one we wish to describe. Thus,
they must have less indices than $\Psi$.

For this case we obtain the following expression for the 
modification of the action due to the redundancy,
\bea{ll}
\delta_G S=\int d^8z&\left\{-2a_1\D_{\a_s}\Psi^{\a(s)\ad(s-1
)}\right.\\
&~~+\left. a_4\Dd_{\ad_s}\bar{\Psi}^{\a(s-1)\ad(s)}\right\}
\D^{\b}\Dd_{\ad_{s-1}}\Lambda_{\b\a(s-1)\ad(s-2)}\\
&+\left\{-a_3\left[\frac{s-1}{s}\right]\Dd_{\ad_s}\D_{\a_{s-1}}
\bar{\Psi}^{\a(s-1)\ad(s)}\right.\\
&~~+\left.\left[-a_3+\frac{s+1}{s}a_4\right]\D_{\a_{s-1}}\Dd_{
\ad_s}\bar{\Psi}^{\a(s-1)\ad(s)}\right\}
\D^{\b}K_{\b\a(s-2)\ad(s-1)}\IEEEyesnumber\\
&+\left\{2a_2\D_{\a_s}\Dd^2\Psi^{\a(s)\ad(s-1)}-a_3\Dd_{
\ad_s}\D^2\bar{\Psi}^{\a(s-1)\ad(s)}\right\}K_{\a(s-1)\ad(s-1)}\\
&+c.c.
\eea
Obviously we can not make all this terms vanish just by 
picking values for the a's without setting them all to zero 
and also we can't introduce compensators with proper 
mass dimensionality and index structure. The way out 
is to give some structure to the gauge parameter $K$. 
So let us choose
\bea{l}
a_1=a_4=0\nonumber\\
\D^{\b}K_{\b\a(s-2)\ad(s-1)}=0\rightarrow K_{\a(s-1)\ad(s-1
)}=\D^{\a_s}L_{\a(s)\ad(s-1)}\IEEEyesnumber\\
2a_2=-a_3\nonumber
\eea
So we find
\bea{ll}
\delta_G S=-a_3\int d^8z
&\D_{\a_s}\Dd^2\Psi^{\a(s)\ad(s-1)}\left(\D^{\b}L_{\b\a(
s-1)\ad(s-1)}+\Dd^{\bd}\bar{L}_{\a(s-1)\bd\ad(s-1)}
\right)\IEEEyesnumber\\
&+c.c.
\eea
This suggests we introduce a real bosonic compensator 
$V_{\a(s-1)\ad(s-1)}$ which transforms like $\delta_G 
V_{\a(s-1)\ad(s-1)}=\D^{\a_s}L_{\a(s)\ad(s-1)}+\Dd^{
\ad_s}\bar{L}_{\a(s-1)\a(s)}$ and couples with the real 
piece of $\D^{\a_s}\Dd^2\Psi_{\a(s)\ad(s-1)}$.

In order to achieve invariance, we add to the action two 
new pieces, a coupling term of $V$ with $\Psi$ and a 
kinetic energy term for $V$. The full action takes the form
\bea{ll}
S=&\int d^8z\ -\frac{1}{2}a_3\Psi^{\a(s)\ad(s-1)}\Dd^2\Psi_{
\a(s)\ad(s-1)}+c.c.\nonumber\\
&+a_3\Psi^{\a(s)\ad(s-1)}\Dd^{\ad_s}\D_{\a_s}\bar{\Psi}_{
\a(s-1)\ad(s)}\nonumber\\
&-a_3 V^{\a(s-1)\ad(s-1)}\D^{\a_s}\Dd^2\Psi_{\a(s)\ad(s-1)
}+c.c.\nonumber\\
&+b_1 V^{\a(s-1)\ad(s-1)}\D^{\g}\Dd^2\D_{\g}V_{\a(s-1)\ad(
s-1)}\IEEEyesnumber\\
&+b_2 V^{\a(s-1)\ad(s-1)}\left\{\D^2,\Dd^2\right\}V_{\a(s-1
)\ad(s-1)}\nonumber\\
&+b_3 V^{\a(s-1)\ad(s-1)}\D_{\a_{s-1}}\Dd^2\D^{\g}V_{\g
\a(s-2)\ad(s-1)}+c.c.\nonumber\\
&+b_4 V^{\a(s-1)\ad(s-1)}\D_{\a_{s-1}}\Dd_{\ad_{s-1}}\D^{
\g}\Dd^{\gd}V_{\g\a(s-2)\gd\ad(s-2)}+c.c.\nonumber
\eea
and it has to be invariant under
\bea{ll}
\delta_G\Psi_{\a(s)\ad(s-1)}&=-\D^2 L_{\a(s)\ad(s-1)}+\left[
\frac{1}{(s-1)!}\right]\Dd_{(\ad_{s-1}}\Lambda_{\a(s)\ad(s-2
))}\IEEEyessubnumber\\
\delta_G V_{\a(s-1)\ad(s-1)}&=\D^{\a_s}L_{\a(s)\ad(s-1)}+
\Dd^{\ad_s}\bar{L}_{\a(s-1)\ad(s)}\IEEEyessubnumber
\eea

The equations of motion of the superfields are the variation 
of the action with respect to the corresponding superfield
\bea{l}
T_{\a(s)\ad(s-1)}=\frac{\delta S}{\delta \Psi^{\a(s)\ad(s-1
)}}\IEEEyessubnumber\\
G_{\a(s-1)\ad(s-1)}=\frac{\delta S}{\delta V^{\a(s-1)\ad(s-1
)}}\IEEEyessubnumber
\eea
and the invariance of the action gives the following Bianchi 
Identities
\bea{l}
\D^2T_{\a(s)\ad(s-1)}+\frac{1}{s!}\D_{(\a_s}G_{\a(s-1))\ad(s-1
)}=0\IEEEyessubnumber\\
\Dd^{\ad_{s-1}}T_{a(s)\ad(s-1)}=0\IEEEyessubnumber
\eea
The satisfaction of the Bianchi identities fix all the coefficients
\bea{lr}
b_1=\frac{1}{2}a_3 ~~ & b_3=0\\
b_2=0 & b_4=0
\eea
and the action takes the form\footnote{Here $c$ is an overall 
unconstrained parameter which can be absorbed into the 
definition of $\Psi$. \newline $~~\,~~~$ We leave it as it is 
for now and fix it later in the component discussion.}
\bea{ll}
S=\int d^8z&\left\{-\frac{1}{2}c\Psi^{\a(s)\ad(s-1)}\Dd^2
\Psi_{\a(s)\ad(s-1)}+c.c.\right.\\
&~~+c\Psi^{\a(s)\ad(s-1)}\Dd^{\ad_s}\D_{\a_s}\bar{\Psi
}_{\a(s-1)\ad(s)}\\
&~~-c V^{\a(s-1)\ad(s-1)}\D^{\a_s}\Dd^2\Psi_{\a(s)\ad(s-1
)}+c.c.\IEEEyesnumber\\
&~~\left.+\frac{1}{2}c V^{\a(s-1)\ad(s-1)}\D^{\g}\Dd^2\D_{
\g}V_{\a(s-1)\ad(s-1)}\right\}
\eea

The equations of motion are
\bea{ll}
T_{\a(s)\ad(s-1)}&=-c\Dd^2\Psi_{\a(s)\ad(s-1)}+\frac{c}{s!}
\Dd^{\ad_s}\D_{(\a_s}\bar{\Psi}_{\a(s-1))\ad(s)}\\
&~~+\frac{c}{s!}\Dd^2\D_{(\a_s}V_{\a(s-1))\ad(s-1)}
\IEEEyessubnumber\\
G_{\a(s-1)\ad(s-1)}&=-c\left(\D^{\a_s}\Dd^2\Psi_{\a(s
)\ad(s-1)}+\Dd^{\ad_s}\D^2\bar{\Psi}_{\a(s-1)\a(s)}\right)\\
&~~+c\D^\g\Dd^2\D_\g V_{\a(s-1)\ad(s-1)}\IEEEyessubnumber
\eea

This is exactly the longitudinal-linear theory presented in 
\cite{Off2} if we solve the constraint superfield and express 
their action in terms of the prepotential.  Now, however we 
gain a different understanding of why the action has to be 
expressed in terms of a superfield like $\Psi$ and why it 
has a gauge transformation as it does. 

The work in \cite{Off2} presented a second theory for integer 
super-helicity, the transverse-linear theory. That theory is
most certainly consistent classically, but violates one of our
assumptions in that some of its auxiliary fields possess
spins greater than that carried by the gauge superfield.
To our knowledge, no studies of the quantum behavior
of these off-shell supersymmetrical and even free theories
has been carried out.  If is our suspicion that the presence
of auxiliary superfields with a higher superspin than the
main gauge superpotential is likely to have a more complicated
ghost structure.  It would be a very interesting investigation
to test this idea.

We have managed to find a superspace action which is 
gauged invariant but still we haven't proved that this theory 
describes an integer super-helicity system. To do so, we 
must show that there is an object like $F_{\a(2s)}$, it is 
chiral and on-shell it satisfies the required by representation 
theory constraints .

Using the equations of motion we can now prove that a chiral 
superfield $F_{\a(2s)}$ exists and satisfies following Bianchi identity:
\bea{ll}
\Dd^{\ad_{2s}}\bar{F}_{\ad(2s)}=&-\frac{i}{(2s-1)!c}\pa^{\a_s}
{}_{(\ad_{2s-1}}\ldots\pa^{\a_1}{}_{\ad_{s}}T_{\a(s)\ad(s-1))}\\
&+\frac{B}{(2s-1)!}\Dd^2\pa^{\a_{s-1}}{}_{(\ad_{2s-1}}\ldots\pa^{
\a_1}{}_{(\ad_{s+1}}
\bar{T}_{\a(s-1)\ad(s))}\IEEEyesnumber\\
&+\frac{1+2cB}{(2s-1)!2c}\Dd_{(\ad_{2s-1}}\pa^{\a_{s-1}}{}_{
\ad_{2s-2}}\ldots\pa^{\a_{1}}{}_{\ad_{s}}G_{\a(s-1)\ad(s-1))}\\
&+\frac{1}{(2s-1)!2c}\Dd_{(\ad_{2s-1}}\D^{\a_s}\pa^{\a_{s-1}}
{}_{\ad_{2s-2}}\ldots\pa^{\a_1}{}_{\ad_{s}}T_{\a(s)\ad(s-1)}
\eea
where
\bea{l}
\bar{F}_{\ad(2s)}=\frac{1}{(2s)!}\D^2\Dd_{(\ad_{2s}}\pa^{\a_{
s-1}}{}_{\ad_{2s-1}}\dots\pa^{\a_{1}}{}_{\ad_{s+1}}\bar{\Psi}_{
\a(s-1)\ad(s))}
\eea
and that shows that if $T_{\a(s)\ad(s-1)}$ = $G_{\a(s-1)\ad(s-1))}$ = 
0, we obtain the desired constraints to describe a super-helicity $
Y=s$ system, where $B$ is a parameter determined by variations 
and definitions.

Before we start investigating the field spectrum of the above action, 
one more comment needs to be made. This specific action and 
superfield configuration is not unique but the simplest representative 
of a two parameter family of equivalent theories.  To see that we 
can perform redefinitions of the superfields. Dimensionality and 
index structure allow us to make the following redefinition of $\Psi$
\bea{l}
\Psi_{\a(s)\ad(s-1)}\rightarrow\Psi_{\a(s)\ad(s-1)}+\frac{z}{s!}\D_{(
\a_s}V_{\a(s-1))\ad(s-1)}\IEEEyesnumber
\eea
where $z$ is a complex parameter. This operation will generate an 
entire class of actions and transformation laws which all are related 
by the above redefinition.
The action is
\bea{ll}
S&=\int d^8w\left\{-\frac{1}{2}c~\Psi^{\a(s)\ad(s-1)}\Dd^2\Psi_{\a(s
)\ad(s-1)}+c.c.\right.\\
&~~+c~\Psi^{\a(s)\ad(s-1)}\Dd^{\ad_s}\D_{\a_s}\bar{\Psi}_{\a(s-1
)\ad(s)}\\
&~~+c(z+\bar{z}-1)~V^{\a(s-1)\ad(s-1)}\D^{\a_s}\Dd^2\Psi_{\a(s
)\ad(s-1)}+c.c.\\
&~~+c\bar{z}~V^{\a(s-1)\ad(s-1)}\Dd^2\D^{\a_s}\Psi_{\a(s)\ad(s-1
)}+c.c.\\
&~~-\left[\frac{s-1}{s}\right]c\bar{z}~V^{\a(s-1)\ad(s-1)}\Dd_{\ad_{
s-1}}\D^{\b}\Dd^{\bd}\Psi_{\b\a(s-1)\bd\ad(s-2)}+c.c.\\
&~~+\frac{1}{2}c(z+\bar{z}-1)^2~V^{\a(s-1)\ad(s-1)}\D^{\g}\Dd^2
\D_{\g}V_{\a(s-1)\ad(s-1)}\IEEEyesnumber\\
&~~+\left[\frac{1}{s}\right]cz\bar{z}~V^{\a(s-1)\ad(s-1)}\left\{\D^2,
\Dd^2\right\}V_{\a(s-1)\ad(s-1)}\\
&~~+\left[\frac{s-1}{2s}\right]cz(z+2\bar{z}-2)~V^{\a(s-1)\ad(s-1)}
\D_{\a_{s-1}}\Dd^2\D^{\g}V_{\g\a(s-2)\ad(s-1)}+c.c.\\
&~~-\left.\left[\frac{(s-1)^2}{2s^2}\right]cz\bar{z}~ V^{\a(s-1)\ad(s-1
)}\D_{\a_{s-1}}\Dd_{\ad_{s-1}}\D^{\g}\Dd^{\gd}V_{\g\a(s-2)\gd\ad(
s-2)}+c.c.\right\}
\eea
and the transformation laws are
\bea{ll}
\delta_G\Psi_{\a(s)\ad(s-1)}&=\left(z-1\right)\D^2 L_{\a(s)\ad(s-1
)}-\frac{z}{s!}\D_{(\a_s}\Dd^{\ad_s}\bar{L}_{\a(s-1))\ad(s)}\\
&~+\left[\frac{1}{(s-1)!}\right]\Dd_{(\ad_{s-1}}\Lambda_{\a(s)\ad(
s-2))}\IEEEyessubnumber\\
\delta_G V_{\a(s-1)\ad(s-1)}&=\D^{\a_s}L_{\a(s)\ad(s-1)}+
\Dd^{\ad_s}\bar{L}_{\a(s-1)\ad(s)}\IEEEyessubnumber
\eea

\section{Projection and Components}
\label{Projection}
~~Although superspace was developed to describe supersymmetric 
theories in a more efficient, compact and clear way, there are still 
some reasons why we would like to study the off-shell component 
structure of the theory.
\begin{enumerate}
\item There are cases where two theories on-shell describe the 
same physical system. Therefore from the path integral point of 
view the theories are equivalent. Nevertheless the off-shell 
structure of the two theories might be completely different. 
Knowledge of the component formulation of the two theories 
will help us decide if they are different theories with the same 
on-shell description or they are the same theory and there is 
a 1-1 mapping between the two. 
\item The off-shell component structure of a supersymmetric 
theory will give us clues about which theories can be used to 
realize higher $\mathcal{N}$ and higher $D$ representations.   
\end{enumerate}
For these reasons we would like to extract the component field 
content of the above superspace action, the number of degrees 
of freedom involved, their transformation law under supersymmetry 
and their gauge transformations.

Previous discussion to this use the Wess-Zumino and explicit 
$\theta$-expansions.  We propose a different technique that will 
illuminate a more natural way to define the component structure 
and make the entire process of finding the component action and 
SUSY-transformation laws efficiently.

Since we want the auxiliary fields of the final action to be gauge 
invariant it might be smart to define them using objects that are 
already gauge invariant. But the superspace action already 
provides us with two gauge invariant objects, the equations 
of motion\footnote{There is also the superfield strength $F_{
\a(2s)}$ but because of dimensionality reasons we can 
\newline $~~~~~~$
not write the action in terms of it.}:
\bea{ll}
T_{\a(s)\ad_(s-1)}=\frac{\d S}{\d\Psi^{\a(s)\ad(s-1)}} ~,&~\left[
T_{\a(s)\ad(s-1)}\right]=3/2\IEEEyessubnumber\\
G_{\a(s-1)\ad_(s-1)}=\frac{\d S}{\d V^{\a(s-1)\ad(s-1)}} ~,&~
\left[G_{\a(s-1)\ad(s-1)}\right]=2\IEEEyessubnumber\\
{}&~G_{\a(s-1)\ad(s-1)}=\bar{G}_{\a(s-1)\ad(s-1)}
\eea

Because they are gauge invariant, if we expand them to components, 
each one of them will be gauge invariant. Furthermore because they 
vanish on-shell each one of these components will vanish as well. 
So it looks like the ideal place to look for the auxiliary component 
structure.

These superfields satisfy a set of equations that we will discover 
as we go along, but at the top of the list we have the Bianchi 
identities\footnote{The Bianchi identities include the entire 
information about redundancy and therefore 
\newline $~~~~~~$
effectively they make everything that could have been gauged 
away, if we had followed
\newline $~~~~~~$
 the WZ-gauge path, disappear} and their consequences:
\bea{ll}
\D^2T_{\a(s)\ad(s-1)}+\frac{1}{s!}\D_{(\a_s}G_{\a(s-1))\ad(s-1)}
=0\leadsto &\D^2G_{\a(s-1)\ad(s-1)}=0\IEEEyessubnumber\\
{}&\Dd^2G_{\a(s-1)\ad(s-1)}=0\\
\Dd^{\ad_{s-1}}T_{a(s)\ad(s-1)}=0\leadsto\Dd^2T_{a(s)\ad(s-1
)}=0\IEEEyessubnumber
\eea
The results of these are that most of the components in the 
expansion of $T$ and $G$ vanish and we are left with very 
few that we can associate with auxiliary fields. For example, 
the bosonic auxiliary fields (dimensionality $2$) have to be 
related to $\Dd_{(\ad_s}T_{\a(s)\ad(s-1))}|,~D^{\a_s}T_{\a(s
)\ad(s-1)}|,$ $G_{\a(s-1)\ad(s-1)}|$ and the fermionic ones 
($3/2,~5/2$) will have to be related to $T_{\a(s)\ad(s-1)}|$,
\\$\D^2T_{\a(s)\ad(s-1)}|$.  So by just looking at the Bianchi 
identities we find for free the spectrum of the auxiliary fields 
of the action and because they are gauge invariant we can 
do a  straightforward counting of their degrees of freedom. 
For the dynamical fields, we can use the superfield strength 
$F_{\a(2s)}$ to connect them with some components of the 
superfields. Instead we will let the action, the equations of 
motion and their properties to guide us to their definition.

But if the equations of motion are the proper objects to 
define the components and we want to find the component 
action of the theory we must be able to express the action 
in terms of the equations of motion. That can be easily done 
by using the definitions of $T$ and $G$ to rewrite the action 
in the following form
\bea{ll}
S=\int d^8z &\left\{~~\frac{1}{2}\Psi^{\a(s)\ad(s-1)}T_{\a(s)
\ad(s-1)}+c.c.\right.\IEEEyesnumber\\
&\left.~+\frac{1}{2}V^{\a(s-1)\ad(s-1)}G_{\a(s-1)\ad(s-1)}\right\}\\
=\int d^4x &\frac{1}{2}\D^2\Dd^2\left(\Psi^{\a(s)\ad(s-1)}
T_{\a(s)\ad(s-1)}\right)+c.c.\\
&+\frac{1}{2}\D^2\Dd^2\left(V^{\a(s-1)\ad(s-1)}G_{\a(s-1
)\ad(s-1)}\right)
\eea
and now we distribute the covariant derivatives.

\subsection{Fermions}
Let us focus on the fermionic action first. After the distribution 
of $\D$'s and the usage of Bianchi identities we find for the 
fermionic Lagrangian:
\bea{lll}
\mathcal{L}_F&=&\frac{1}{2}\D^2\Dd^2\Psi^{\a(s)\ad(s-1)}|
T_{\a(s)\ad(s-1)}|\\
&+&\frac{1}{2}\left(\Dd^2\Psi^{\a(s)\ad(s-1)}-\frac{1}{s!}
\Dd^2\D^{(\a_s}V^{\a(s-1))\ad(s-1)}\right)|\D^2 T_{\a(s)
\ad(s-1)}|\\
&-&\frac{1}{2}\frac{1}{(s+1)!s!}\D^{(\a_{s+1}}\Dd^{(\ad_s}
\Psi^{\a(s))\ad(s-1))}|\frac{1}{(s+1)!s!}\D_{(\a_{s+1}}
\Dd_{(\ad_s}T_{\a(s))\ad(s-1))}|\\
&+&\frac{1}{2}\frac{s}{s+1}\frac{1}{s!}\D_{\g}\Dd^{(\ad_s}
\Psi^{\g\a(s-1)\ad(s-1))}|\frac{1}{s!}\D^{\a_s}\Dd_{(\ad_s}
T_{\a(s)\ad(s-1))}|\IEEEyesnumber\\
&-&\frac{s-1}{2s}\Dd^2\D_{\g}V^{\g\a(s-2)\ad(s-1)}|\D^{
\a_{s-1}}G_{\a(s-1)\ad(s-1)}|\\
&+&c.c.
\eea
At this point we can show that $T$ and $G$ satisfy a few 
more identities:
\bea{ll}
\frac{1}{(s+1)!s!}&\D_{(\a_{s+1}}\Dd_{(\ad_s}T_{\a(s))
\ad(s-1))}=\\
&=-\frac{ic}{(s+1)!}\pa_{(\a_{s+1}}{}^{\ad_{s+1}}\left[
\frac{1}{(s+1)!s!}\Dd_{(\ad_{s+1}}\D_{(\a_s}\bar{\Psi
}_{\a(s-1)))\ad(s))}\right]\\
&~~+\frac{ic}{(s+1)!s!}\frac{s}{s+1}\pa_{(\a_{s+1}(
\ad_{s}}\left[\frac{1}{s!}\Dd^{\gd}\D_{(\a_s}\bar{\Psi}_{
\a(s-1)))\gd\ad(s-1))}\right]
\eea
%%%%
%%%%
\bea{ll}
\frac{1}{s!}\D^{\a_s}\Dd_{(\ad_s}T_{\a(s)\ad(s-1))}
&=\frac{i}{s!}\frac{s+1}{s}\pa^{\a_s}{}_{(\ad_s}T_{
\a(s)\ad(s-1))}\\
&~~+\frac{s+1}{s}\Dd^2\bar{T}_{\a(s-1)\ad(s)}\\
&~~-\frac{ic}{s!(s+1)!}\pa^{\a_s\ad_{s+1}}\Dd_{
(\ad_{s+1}}\D_{(\a_s}\bar{\Psi}_{\a(s-1))\ad(s))}\\
&~~-\frac{ic}{s!s!}\frac{2s+1}{s(s+1)}\pa^{\a_s}
{}_{(\ad_s}\Dd^{\gd}\D_{(\a_s}\bar{\Psi}_{\a(s-1))
\gd\ad(s-1))}\\
&~~-\frac{ic}{s!s!}\frac{s^2-1}{s}\pa_{(\a_{s-1
}(\ad_s}\Dd^2\D^{\g}V_{\g\a(s-2))\ad(s-1))}
\eea
%%%%
%%%%
\bea{ll}
\D^{\a_{s-1}}G_{\a(s-1)\ad(s-1)}&=i\pa^{\a_{s-1}
\ad_s}\bar{T}_{\a(s-1)\ad(s)}\\
&~~-\frac{ic}{s!}\pa^{\a_{s-1}\ad_s}\D^{\g}\Dd_{
(\ad_s}\Psi_{\g\a(s-1)\ad(s-1))}\\
&~~-ic\frac{s-1}{s!}\pa^{\a_{s-1}}{}_{(\ad_{s-1}}
\D^2\Dd^{\gd}V_{\a(s-1)\gd\ad(s-2))}
\eea
%%%%
%%%%
\bea{ll}
\Dd^2\bar{T}_{\a(s-1)\ad(s)}+\frac{i}{s!}\pa^{
\a_s}{}_{(\ad_s}&T_{\a(s)\ad(s-1))}=\\
&=\frac{ic}{s!s!}\pa^{\a_s}{}_{(\ad_s}\Dd^{\gd}
\D_{(\a_s}\bar{\Psi}_{\a(s-1))\gd\ad(s-1))}\\
&~~-c\Dd^2\D^2\bar{\Psi}_{\a(s-1)\ad(s)}\\
&~~+ic\frac{(s-1)}{s!s!}\pa_{(\a_{s-1}(\ad_{s}}
\Dd^2\D^{\g}V_{\g\a(s-2))\ad(s-1))}
\eea
We notice that in all the above there are some combinations 
that appear repeatedly.  So let us define the following fields:
\bea{l}
\frac{1}{s!(s+1)!}\D_{(\a_{s+1}}\Dd_{(\ad_s}\Psi_{\a(s))\ad(
s-1))}|\equiv N_1 \psi_{\a(s+1)\ad(s)}\\
\frac{1}{s!}\Dd^{\ad_s}\D_{(\a_s}\bar{\Psi}_{\a(s-1))\ad(s)}|
\equiv N_2 \psi_{\a(s)\ad(s-1)}\\
\D^2\Dd^{\ad_{s-1}}V_{\a(s-1)\ad(s-1)}|\equiv N_3\psi_{
\a(s-1)\ad(s-2)}
\eea
where $N_1,~N_2,~N_3,~N_4$ are some overall normalization, 
to be fixed later as needed.

Putting everything together we find the fermionic terms of the 
Lagrangian
\bea{ll}
\mathcal{L}_F=&-\frac{1}{2c}T^{\a(s)\ad(s-1)}|\left(2\D^2T_{
\a(s)\ad(s-1)}+\frac{i}{s!}\pa_{(\a_s}{}^{\ad_s}\bar{T}_{\a(s-1
))\ad(s)}\right)|+c.c.\\
&-ic|N_1|^2~\bar{\psi}^{\a(s)\ad(s+1)}\pa^{\a_{s+1}}{}_{
\ad_{s+1}}\psi_{\a(s+1)\ad(s)}\\
&-ic\frac{s}{s+1}N_1N_2~\psi^{\a(s+1)\ad(s)}\pa_{\a_{s+1}
\ad_s}\psi_{\a(s)\ad(s-1)}+c.c.\\
&+ic\frac{2s+1}{(s+1)^2}|N_2|^2~\bar{\psi}^{\a(s-1)\ad(s)}
\pa^{\a_{s}}{}_{\ad_{s}}\psi_{\a(s)\ad(s-1)}\\
&+ic\frac{s-1}{s}N_2N_3~\psi^{\a(s)\ad(s-1)}\pa_{\a_{s}
\ad_{s-1}}\psi_{\a(s-1)\ad(s-2)}+c.c.\\
&+ic\left(\frac{s-1}{s}\right)^2|N_3|^2~\bar{\psi}^{\a(s-2)
\ad(s-1)}\pa^{\a_{s-1}}{}_{\ad_{s-1}}\psi_{\a(s-1)\ad(s-2)}\\
\eea
The first term in the Lagrangian is the algebraic term of 
two auxiliary fields and the rest of the terms have exactly 
the structure of a theory that describes helicity $h=s+1/2
$\cite{BK}\footnote{We are following the conventions of 
\cite{GGRS} which differ from the conventions used in 
\cite{BK}.}. For an exact match we choose coefficients
\bea{ll}
c=-1~,~& N_2=1\\
N_1=1~,~& N_3=-\frac{s}{s-1}
\eea
So the fields that appear in the fermionic action are defined 
as:
\bea{l}
\rho_{\a(s)\ad(s-1)}\equiv T_{\a(s)\ad(s-1)}|\\
\b_{\a(s)\ad(s-1)}\equiv\D^2T_{\a(s)\ad(s-1)}|+\frac{i}{2s!}
\pa_{(\a_s}{}^{\ad_s}\bar{T}_{\a(s-1))\ad(s)}|\\
\psi_{a(s+1)\ad(s)}\equiv\frac{1}{s!(s+1)!}\D_{(\a_{s+1
}}\Dd_{(\ad_s}\Psi_{\a(s))\ad(s-1))}|\IEEEyesnumber\\
\psi_{\a(s)\ad(s-1)}\equiv\frac{1}{s!}\Dd^{\ad_s}\D_{(
\a_s}\bar{\Psi}_{\a(s-1))\ad(s)}|\\
\psi_{\a(s-1)\ad(s-2)}\equiv-\frac{s-1}{s}\D^2\Dd^{\ad_{
s-1}}V_{\a(s-1)\ad(s-1)}|
\eea
The Lagrangian is
\bea{ll}
\mathcal{L}_F=&\rho^{\a(s)\ad(s-1)}\beta_{\a(s)\ad(s-1
)}+c.c.\\
&+i~\bar{\psi}^{\a(s)\ad(s+1)}\pa^{\a_{s+1}}{}_{\ad_{s+1
}}\psi_{\a(s+1)\ad(s)}\\
&+i\left[\frac{s}{s+1}\right]~\psi^{\a(s+1)\ad(s)}\pa_{\a_{
s+1}\ad_s}\psi_{\a(s)\ad(s-1)}+c.c.\\
&-i\left[\frac{2s+1}{(s+1)^2}\right]~\bar{\psi}^{\a(s-1)\ad(
s)}\pa^{\a_{s}}{}_{\ad_{s}}\psi_{\a(s)\ad(s-1)}\IEEEyesnumber\\
&+i~\psi^{\a(s)\ad(s-1)}\pa_{\a_{s}\ad_{s-1}}\psi_{\a(s-1
)\ad(s-2)}+c.c.\\
&-i~\bar{\psi}^{\a(s-2)\ad(s-1)}\pa^{\a_{s-1}}{}_{\ad_{s-1
}}\psi_{\a(s-1)\ad(s-2)}\\
\eea
and the gauge transformations of the fields are
\bea{ll}
\d_G\rho_{\a(s)\ad(s-1)}=0~,~&\d_G\psi_{\a(s+1)\ad(s)}
=\frac{1}{s!(s+1)!}\pa_{(\a_{s+1}(\ad_s}\xi_{\a(s))\ad(s-1))}\\
\d_G\b_{\a(s)\ad(s-1)}=0~,~&\d_G\psi_{\a(s)\ad(s-1)}=
-\frac{1}{s!}\pa_{(\a_{s}}{}^{\ad_s}\bar{\xi}_{\a(s-1))\ad(s
)}\IEEEyesnumber\\
&\d_G\psi_{\a(s-1)\ad(s-2)}=\frac{s-1}{s}\pa^{\a_s\ad_{
s-1}}\xi_{\a(s)\ad(s-1)}\\
& \text{with } \xi_{\a(s)\ad(s-1)}=-i\D^2L_{\a(s)\ad(s-1)}|
\eea
\subsection{Bosons}
~~~
For the bosonic action we follow exactly the same procedure 
as was presented for the fermionic sector. The fields that 
appear in the action are defined as:
\bea{l}
U_{\a(s+1)\ad(s-1)}\equiv\frac{1}{(s+1)!}\D_{(\a(s+1)}T_{\a(
s))\ad(s-1)}|\\
u_{\a(s)\ad(s)}\equiv\frac{1}{2s!}\left\{\Dd_{(\ad_s}T_{\a(s)
\ad(s-1))}-\D_{(\a_s}\bar{T}_{\a(s-1))\ad(s)}\right\}|\\
v_{\a(s)\ad(s)}\equiv -\frac{i}{2s!}\left\{\Dd_{(\ad_s}T_{\a(s)
\ad(s-1))}+\D_{(\a_s}\bar{T}_{\a(s-1))\ad(s)}\right\}|\\
A_{\a(s-1)\ad(s-1)}\equiv G_{\a(s-1)\ad(s-1)}|-\frac{s}{2s+1
}\left(\D^{\a_s}T_{\a(s)\ad(s-1)}+\Dd^{\ad_s}\bar{T}_{\a(s)
\ad(s-1)}\right)|\\
S_{\a(s-1)\ad(s-1)}\equiv\frac{1}{2}\left\{\D^{\a_s}T_{\a(s)
\ad(s-1)}+\Dd^{\ad_s}\bar{T}_{\a(s)\ad(s-1)}\right\}
|\IEEEyesnumber\\
P_{\a(s-1)\ad(s-1)}\equiv -\frac{i}{2}\left\{\D^{\a_s}T_{\a(
s)\ad(s-1)}-\Dd^{\ad_s}\bar{T}_{\a(s)\ad(s-1)}\right\}|\\
h_{\a(s)\ad(s)}\equiv\frac{1}{\sqrt{2}}\left\{\frac{1}{s!}\D_{
(\a_s}\bar{\Psi}_{\a(s-1))\ad(s)}-\frac{1}{s!}\Dd_{(\ad_s}
\Psi_{\a(s)\ad(s-1))}\right.\\
~~~~~~~~~~~~~~~~~~~~~-\left.\frac{1}{2s!s!}\left[\D_{
(\a_s},\Dd_{(\ad_s}\right]V_{\a(s-1))\ad(s-1))}\right\}|\\
h_{\a(s-2)\ad(s-2)}\equiv -\frac{1}{2\sqrt{2}}\frac{s-1}{
s^2}\left[\D^{\a_{s-1}},\Dd^{\ad_{s-1}}\right]V_{\a(s-1)\ad(
s-1)}|
\eea
the gauge transformations are
\bea{ll}
\d_G U_{\a(s+1)\ad(s-1)}=0,~~&\d_G A_{\a(s-1)\ad(s-1
)}=0\\
\d_G u_{\a(s)\ad(s)}=0,~~&\d_G S_{\a(s-1)\ad(s-1)}=0
\IEEEyesnumber\\
\d_G v_{\a(s)\ad(s)}=0,~~&\d_G P_{\a(s-1)\ad(s-1)}=0\\
\d_G h_{\a(s)\ad(s)}=\frac{1}{s!s!}\pa_{(\a_s(\ad_s}\zeta_{
\a(s-1))\ad(s-1))}\\
\d_G h_{\a(s-2)\ad(s-2)}=\frac{s-1}{s^2}\pa^{\a_{s-1}\ad_{
s-1}}\zeta_{\a(s-1)\ad(s-1)}\\
\eea
where
\bea{l}
\zeta_{\a(s-1)\ad(s-1)}=\frac{i}{2\sqrt{2}}\left(\D^{\a_s}L_{
\a(s)\ad(s-1)}-\Dd^{\ad_s}\bar{L}_{\a(s-1)\ad(s)}\right)
\eea
and the Lagrangian is
\bea{ll}
\mathcal{L}_B=&-\frac{1}{2}~U^{\a(s+1)\ad(s-1)}U_{\a(s+1
)\ad(s-1)}+c.c.\\
&+u^{\a(s)\ad(s)}u_{\a(s)\ad(s)}\\
&+v^{\a(s)\ad(s)}v_{\a(s)\ad(s)}\\
&-\left[\frac{2s+1}{4s}\right]A^{\a(s-1)\ad(s-1)}A_{\a(s-1)
\ad(s-1)}\\
&-\left[\frac{s^2}{(2s+1)(s+1)}\right]S^{\a(s-1)\ad(s-1)}S_{
\a(s-1)\ad(s-1)}\\
&-\left[\frac{s^2}{s+1}\right]P^{\a(s-1)\ad(s-1)}P_{\a(s-1)
\ad(s-1)}\\
&+h^{\a(s)\ad(s)}\Box h_{\a(s)\ad(s)}\\
&-\frac{s}{2}~h^{\a(s)\ad(s)}\pa_{\a_s\ad_s}\pa^{\g\gd}
h_{\g\a(s-1)\gd\ad(s-1)}\\
&+s(s-1)~h^{\a(s)\ad(s)}\pa_{\a_s\ad_s}\pa_{\a_{s-1}
\ad_{s-1}}h_{\a(s-2)\ad(s-2)}\\
&-s(2s-1)~h^{\a(s-2)\ad(s-2)}\Box h_{\a(s-2)\ad(s-2)}\\
&-\left[\frac{s(s-2)^2}{2}\right]h^{\a(s-2)\ad(s-2)}\pa_{
\a_{s-2}\ad_{s-2}}\pa^{\g\gd}h_{\g\a(s-3)\gd\ad(s-3)}
\eea

\subsection{Off-shell degrees of freedom}
Let us count the bosonic degrees of freedom of the theory:
\begin{center}
\begin{tabular}{|c|c|c|c|}
\hline 
\emph{fields} & \emph{d.o.f} & \emph{redundancy} & 
\emph{net} \\ 
\hline 
$h_{\a(s)\ad(s)}$ & $(s+1)^2$ & \multirow{2}{*}{$s^2$} 
& \multirow{2}{*}{$s^2+2$} \\ 
\cline{1-2}
$h_{\a(s-2)\ad(s-2)}$ & $(s-1)^2$ & & \\ 
\hline 
$u_{\a(s)\ad(s)}$ & $(s+1)^2$ & 0 & $(s+1)^2$  \\ 
\hline 
$v_{\a(s)\ad(s)}$ & $(s+1)^2$ & 0 & $(s+1)^2$ \\ 
\hline 
$A_{\a(s-1)\ad(s-1)}$ & $s^2$ & 0 & $s^2$ \\ 
\hline 
$U_{\a(s+1)\ad(s-1)}$ & $2(s+2)s$ & 0 & $2(s+2)s$ \\ 
\hline 
$S_{\a(s-1)\ad(s-1)}$ & $s^2$ & 0 & $s^2$ \\ 
\hline 
$P_{\a(s-1)\ad(s-1)}$ & $s^2$ & 0 & $s^2$ \\ 
\hline\hline 
\multicolumn{2}{c|}{} & \emph{Total} & $8s^2+8s+4$ \\ 
\cline{3-4}
\end{tabular}
\end{center}
and the same counting for the Fermionic degrees of 
freedom:
\begin{center}
\begin{tabular}{|c|c|c|c|}
\hline 
\emph{fields} & \emph{d.o.f} & \emph{redundancy} 
& \emph{net} \\ 
\hline 
$\psi_{\a(s+1)\ad(s)}$ & $2(s+2)(s+1)$ & \multirow{
3}{*}{$2(s+1)s$} & \multirow{3}{*}{$4s^2+4s+4$} \\ 
\cline{1-2}
$\psi_{\a(s)\ad(s-1)}$ & $2(s+1)s$ & & \\ 
\cline{1-2}
$\psi_{\a(s-1)\ad(s-2)}$ & $2s(s-1)$ & & \\ 
\hline 
$\rho_{\a(s)\ad(s-1)}$ & $2(s+1)s$ & 0 & $2(s+1)s$ \\ 
\hline 
$\b_{\a(s)\ad(s-1)}$ & $2(s+1)s$ & 0 & $2(s+1)s$ \\ 
\hline\hline 
\multicolumn{2}{c|}{} & \emph{Total} & $8s^2+8s+4$ \\ 
\cline{3-4}
\end{tabular}
\end{center}

\subsection{SUSY-transformation laws}
The last thing left to do is to find explicit expressions for the 
SUSY-transformation laws of the fields. The transformation 
under susy can be easily calculated by the action of the 
SUSY-generators on the specific component. In terms of 
the covariant derivatives $\D(\Dd)$ we 
see that
\be
\d_S\text{Component}=-\left(\e^{\b}\D_{\b}+\ed^{\bd}
\Dd_{\bd}\right)\text{Component}|\nonumber
\ee

But not all the fields are on equal footing. The dynamical ones 
($\in \mathcal{D}$) are treated as equivalence classes, in 
other words they have a gauge transformation of the form 
$\{\mathcal{D}\}\sim\{\mathcal{D}\}+\pa\left(\mathcal{\zeta}\right)$. 
Hence when we apply the susy transformation they will possess 
an extra term in the gauge parameter space
\bea{c}
\d_S\{\mathcal{D}\}\sim\d_S\{\mathcal{D}\}+\pa\left(\d_S
\mathcal{\zeta}\right)
\eea
This says that we must identify these two classes as well, 
therefore we can ignore any terms in the transformation law 
of the dynamical fields that have the same structure as their 
gauge transformation.

With all that in mind we find for the transformation of the fermionic 
fields:
\bea{ll}
\d_S\rho_{\a(s)\ad(s-1)}=&-\e^{\a_{s+1}}U_{\a(s+1)\ad(s-1
)}\\
&+\frac{s}{(s+1)!}\e_{(\a_s}\left[S_{\a(s-1))\ad(s-1)}+iP_{
\a(s-1))\ad(s-1)}\right]\IEEEyesnumber\\
&-\ed^{\ad_s}\left[u_{\a(s)\ad(s)}+iv_{\a(s)\ad(s)}\right]
\eea
\bea{ll}
\d_S\b_{\a(s)\ad(s-1)}=&-i\ed^{\bd}\pa^{\a_{s+1}}{}_{\bd}
U_{\a(s+1)\ad(s-1)}\\
&-\frac{i}{2s!}\ed^{\ad_{s+1}}\pa_{(\a_s}{}^{\ad_s}\bar{
U}_{\a(s-1))\ad(s+1)}\\
&+\frac{i}{2s!}\e^{\b}\pa_{(\a_s}{}^{\ad_s}\left[u_{\b\a(s-1
))\ad(s)}-iv_{\b\a(s-1))\ad(s)}\right]\\
&+\frac{i}{2}\frac{1}{s!s!}\ed^{\ad_s}\pa_{(\a_{s}(\ad_{s
}}A_{\a(s-1))\ad(s-1))}\\
&+\frac{i}{2}\left[\frac{2s^2-1}{(s+1)(2s+1)}\right]\frac{1}{s!s!}
\ed^{\ad_s}\pa_{(\a_s(\ad_s}S_{\a(s-1))\ad(s-1))}\\
&+\frac{1}{2}\left[\frac{2s^2-2s-1}{s+1}\right]\frac{1}{s!s!}
\ed^{\ad_s}\pa_{(\a_s(\ad_s}P_{\a(s-1))\ad(s-1))}
\IEEEyesnumber\\
&-\frac{i}{2}\left[\frac{(s-1)^2}{s(s+1)}\right]\frac{1}{s!(s-1
)!}\ed_{(\ad_{s-1}}\pa_{(\a_s}{}^{\gd}S_{\a(s-1))\gd\ad(s-2))}\\
&+\frac{1}{2}\left[\frac{(s-1)(3s+1)}{s(s+1)}\right]\frac{1}{s!(s-1)!}
\ed_{(\ad_{s-1}}\pa_{(\a_s}{}^{\gd}P_{\a(s-1))\gd\ad(s-2))}\\
&-\sqrt{2}\ed^{\ad_s}\Box h_{\a(s)\ad(s)}\\
&+\frac{s}{\sqrt{2}}\frac{1}{s!s!}\ed^{\ad_s}\pa_{(\a_s(\ad_s}
\pa^{\g\gd}h_{\g\a(s-1))\gd\ad(s-1))}\\
&-\frac{s(s-1)}{\sqrt{2}}\frac{1}{s!s!}\ed^{\ad_s}\pa_{(\a_s(\ad_s
}\pa_{\a_{s-1}\ad_{s-1}}h_{\a(s-2))\ad(s-2))}
\eea
\bea{ll}
\d_S\psi_{\a(s+1)\ad(s)}=&-\frac{1}{s!}\ed_{(\ad_s}U_{\a(s+1
)\ad(s-1))}\\
&-\frac{1}{(s+1)!}\e_{(\a_{s+1}}\left[u_{\a(s))\ad(s)}-iv_{\a(s
))\ad(s)}\right]\IEEEyesnumber\\
&+\frac{i\sqrt{2}}{(s+1)!}\ed^{\bd}\pa_{(\a_{s+1}\bd}h_{\a(s))\ad(s)}
\eea
\bea{ll}
\d_S\psi_{\a(s)\ad(s-1)}=&\ed^{\ad_s}\left[u_{\a(s)\ad(s)}+iv_{
\a(s)\ad(s)}\right]\\
&-\frac{1}{s!}\frac{s}{2s+1}\e_{(\a_s}S_{\a(s-1))\ad(s-1)}\\
&-\frac{is}{s!}\e_{(\a_s}P_{\a(s-1))\ad_(s-1)}\\
&+\frac{1}{s!}\frac{s+1}{2s}\e_{(\a_s}A_{\a(s-1))\ad(s-1)}
\IEEEyesnumber\\
&+i\frac{s-1}{\sqrt{2}}\e^{\b}\pa_{\b}{}^{\ad_s}h_{\a(s)\ad(s)}\\
&+i\frac{(s+1)s(s-1)}{\sqrt{2}s!s!}\e_{(\a_s}\pa_{\a_{s-1}(\ad_{
s-1}}h_{\a(s-2))\ad(s-2))}\\
\eea
\bea{ll}
\d_S\psi_{\a(s-1)\ad(s-2)}=&\frac{1}{2}\frac{(s-1)(2s+1)}{s^2}
\ed^{\ad_{s-1}}A_{\a(s-1)\ad(s-1)}\IEEEyesnumber\\
&+\frac{i}{\sqrt{2}}\frac{(s-1)^2}{s}\frac{1}{(s-1)!^2}\ed^{\ad_{s-1}}
\pa_{(\a_{s-1}(\ad_{s-1}}h_{\a(s-2))\ad(s-2))}\\
&-i\sqrt{2}\frac{(s-1)^2}{s}\frac{1}{(s-1)!^2}\pa_{(\a_{s-1}}{}^{
\ad_{s-1}}\ed_{(\ad_{s-1}}h_{\a(s-2))\ad(s-2))}\\
\eea
The SUSY-transformation laws for the bosonic fields are:
\bea{ll}
\d_S U_{\a(s+1)\ad(s-1)}=&\frac{1}{(s+1)!}\e_{(\a_{s+1}}\b_{\a(s))
\ad(s-1)}\\
&-\frac{i}{2}\frac{1}{(s+1)!}\e_{(\a_{s+1}}\pa_{\a_s}{}^{\ad_s}
\bar{\rho}_{\a(s-1))\ad(s)}\\
&-\frac{i}{(s+1)!}\ed^{\bd}\pa_{(\a_{s+1}\bd}\rho_{\a(s))\ad(s-1)}
\IEEEyesnumber\\
&-\frac{i}{(s+1)!}\ed^{\ad_s}\pa_{(\a_{s+1}}{}^{\ad_{s+1}}\bar{
\psi}_{\a(s))\ad(s+1)}\\
&-i\frac{s}{s+1}\frac{1}{(s+1)!s!}\ed^{\ad_s}\pa_{(\a_{s+1}(\ad_s
}\psi_{\a(s))\ad(s-1))}\\
\eea
\bea{ll}
\d_S\left(u_{\a(s)\ad(s)}+iv_{\a(s)\ad(s)}\right)=&\frac{i}{(s+1)!}
\e^{\a_{s+1}}\pa_{(\a_{s+1}}{}^{\ad_{s+1}}\bar{\psi}_{\a(s))\ad(
s+1)}\\
&-i\frac{s}{s+1}\frac{1}{s!}\e_{(\a_s}\pa^{\g\ad_{s+1}}\bar{
\psi}_{\g\a(s-1))\ad(s+1)}\\
&+i\frac{s}{s+1}\frac{1}{(s+1)!s!}\e^{\a_{s+1}}\pa_{(\a_{s+1}(
\ad_s}\psi_{\a(s))\ad(s-1))}\\
&+i\frac{2s+1}{(s+1)^2}\frac{1}{s!s!}\e_{(\a_s}\pa^{\g}{}_{(\ad_s
}\psi_{\g\a(s-1))\ad(s-1))}\IEEEyesnumber\\
&+i\frac{1}{s!s!}\e_{(\a_s}\pa_{\a_{s-1}(\ad_{s}}\bar{\psi}_{\a(
s-2))\ad(s-1))}\\
&+\frac{1}{s!}\e_{(\a_s}\bar{\b}_{\a(s-1))\ad(s)}\\
&+\frac{i}{2}\frac{1}{s!s!}\e_{(\a_s}\pa^{\g}{}_{(\ad_s}\rho_{\g\a(
s-1))\ad(s-1))}
\eea
\bea{ll}
\d_S A_{\a(s-1)\ad(s-1)}=&-\frac{i}{2s+1}\frac{1}{s!}\ed^{\ad_s}
\pa^{\a_s}{}_{(\ad_s}\rho_{\a(s)\ad(s-1)}+c.c.\\
&+i\frac{(s-1)(s+1)}{s(2s+1)}\frac{1}{(s-1)!}\ed_{(\ad_{s-1}}\pa^{
\a_s\gd}\rho_{\a(s)\gd\ad(s-2))}+c.c.\\
&+i\frac{s}{2s+1}\ed^{\ad_s}\pa^{\a_s\ad_{s+1}}\bar{\psi}_{\a(
s)\ad(s+1)}+c.c.\\
&-\frac{i}{s+1}\frac{1}{s!}\ed^{\ad_s}\pa^{\a_s}{}_{(\ad_s}\psi_{
\a(s)\ad(s-1))}+c.c.\IEEEyesnumber\\
&+i\frac{s-1}{s!}\e_{(\a_{s-1}}\pa^{\g\ad_s}\bar{\psi}_{\g\a(s-2))
\ad(s)}+c.c.\\
&+i\frac{s+1}{2s+1}\frac{1}{(s-1)!s!}\ed^{\ad_s}\pa_{(\a_{s-1}
(\ad_{s}}\bar{\psi}_{\a(s-2))\ad(s-1))}+c.c.\\
&-i\frac{s-1}{s!(s-1)!}\e_{(\a_{s-1}}\pa^{\g}{}_{(\ad_{s-1}}\psi_{
\g\a(s-2))\ad(s-2))}+c.c.\\
\eea
\bea{ll}
\d_S\left(S_{\a(s-1)\ad(s-1)}\right. &\left.+iP_{\a(s-1)\ad(s-1)}
\right)=\\
&=\e^{\a_s}\b_{\a(s)\ad(s-1)}\\
&~~+\frac{s+1}{s}\ed^{\ad_s}\bar{\b}_{\a(s-1)\ad(s)}\\
&~~-\frac{i}{2s!}\e^{\a_s}\pa_{(\a_s}{}^{\ad_s}\bar{\rho}_{\a(s-1)
\ad(s)}\\
&~~-i\frac{s-1}{2s}\frac{1}{s!}\ed^{\ad_s}\pa^{\a_s}{}_{(\ad_s}
\rho_{\a(s)\ad(s-1))}\IEEEyesnumber\\
&~~+i\frac{s-1}{s!}\ed_{(\ad_{s-1}}\pa^{\a_s\gd}\rho_{\a(s)\gd
\ad(s-2))}\\
&~~-i\ed^{\ad_s}\pa^{\a_s\ad_{s+1}}\bar{\psi}_{\a(s)\ad(s+1)}\\
&~~+i\frac{2s+1}{s(s+1)}\frac{1}{s!}\ed^{\ad_s}\pa^{\a_s}{}_{(
\ad_s}\psi_{\a(s)\ad(s-1)}\\
&~~+i\frac{s+1}{s}\frac{1}{(s-1)!s!}\ed^{\ad_s}\pa_{(\a_{s-1}(
\ad_s}\bar{\psi}_{\a(s-2))\ad(s-1))}
\eea
\bea{ll}
\d_S h_{\a(s)\ad(s)}=&\frac{1}{\sqrt{2}s!}\e_{(\a_s}\bar{\rho}_{\a(
s-1))\ad(s)}+c.c.\\
&+\frac{1}{\sqrt{2}}\ed^{\ad_{s+1}}\bar{\psi}_{\a(s)\ad(s+1)}+c.c.
\IEEEyesnumber\\
&-\frac{1}{\sqrt{2}(s+1)}\frac{1}{s!}\ed_{(\ad_s}\psi_{\a(s)\ad(s-1
))}+c.c.
\eea
\bea{l}
\d_S h_{\a(s-2)\ad(s-2)}=-\frac{1}{\sqrt{2}s}\e^{\a_{s-1}}\psi_{\a(
s-1)\ad(s-2)}+c.c.\IEEEyesnumber
\eea
\section{Summary}
~~ We started with a quick review of the representation theory of 
the little group of the Super-Poincar\'{e} group and then we required 
the massless limit of an irreducible massive superspin $Y$ representation 
give us the massless irreducible representation with the same value of 
super-helicity. This forced us to promote the fields used to build the 
theory to equivalence classes and introduce a redundancy. The 
invariance of the physical degrees of freedom of the theory under 
this redundancy fully determines the action of the theory. In this way 
we reproduce the arbitrary integer super-helicity theory but in terms 
of the prepotentials. We recognized that this action is a member of a 
larger two parameter family of equivalent actions, all of which
are connected through superfield redefinitions.

Then we focussed on the off-shell component structure of this superspace 
theory. We presented an alternative technique of defining the field content 
of the theory, using the equations of motion and their Bianchi identities, 
which encode all the information about invariance. Finally we applied it to 
the derivation of the component action and the SUSY-transformation laws 
of the fields involved.

\section*{Acknowledgements}
~~ K. Koutrolikos wants to thank Dr. W.D. Linch and Dr. K. Stiffler 
for useful comments and discussions. This research has been 
supported in part by NSF Grant  PHY-09-68854, the J.~S. Toll 
Professorship endowment and the UMCP Center for String \& 
Particle Theory.

\end{document}